\documentclass{article}
\usepackage{times}
\usepackage[onecolumn]{emulateapj}
\usepackage{natbib}

\usepackage{flushrt}
\def\apj{{ApJ}}                 
\def\apjl{{ApJ}}                
\def\apjs{{ApJS}}               
\def\aap{{A\&A}}                
\def\mnras{{MNRAS}}             
\def\prd{{Phys.~Rev.~D}}        

\newcommand{\ltsima}{$\; \buildrel < \over \sim \;$}
\newcommand{\lsim}{\lower.5ex\hbox{\ltsima}}
\newcommand{\gtsima}{$\; \buildrel > \over \sim \;$}
\newcommand{\gsim}{\lower.5ex\hbox{\gtsima}}

\def\gtrsim{\mathrel{\hbox{\rlap{\hbox{\lower4pt\hbox{$\sim$}}}\hbox{$>$}}}}

\def\lesssim{\mathrel{\hbox{\rlap{\hbox{\lower4pt\hbox{$\sim$}}}\hbox{$<$}}}}

\begin{document}
\title{Three-point Statistics From A New Perspective}

\author{Istv\'an Szapudi \footnote{ Institute for Astronomy, 
University of Hawaii, 2680 
Woodlawn Dr, Honolulu, HI 96822} 
}

\begin{abstract}

Multipole expansion of spatial three-point statistics
is introduced as a tool for investigating and displaying configuration 
dependence. The novel  parametrization
renders the relation between bi-spectrum and three-point correlation function 
especially transparent as a set of two-dimensional Hankel transforms.
It is expected on theoretical grounds, that three-point statistics
can be described accurately with only a few multipoles. In particular,
we show that in the weakly non-linear regime, the multipoles
of the reduced bispectrum, $Q_l$, are significant only up to quadrupole.
Moreover,  the non-linear bias in the weakly non-linear regime only affects 
the monopole order of these statistics. As a consequence,
a simple, novel set of estimators can be constructed 
to constrain galaxy bias.
In addition,  the quadrupole to dipole ratio is independent of the bias,
thus it becomes a novel diagnostic of the underlying theoretical
assumptions: weakly non-linear gravity and perturbative local bias. To
illustrate the use of our approach, we
present predictions based on both power law, and CDM models.
We show that the presently favoured SDSS-WMAP concordance model displays strong
``baryon bumps'' in the $Q_l$'s. Finally, we sketch out three
practical techniques estimate these novel quantities: 
they amount to new, and for the first time edge corrected,
estimators for the bispectrum.

\end{abstract}

\section{Introduction}

Three-point statistics are on track to become main stream tools
in astronomy. They have been used by several authors with considerable
success to constrain the statistical bias between the distribution
of galaxies and dark matter 
\citep[e.g.][]{Fry1994,JingBoerner1998,FriemanGaztanaga1999,
SzapudiEtal2000,ScoccimarroEtal2001,VerdeEtal2002}, 
as well as to constrain primordial
non-Gaussianity of the Cosmic Microwave Background \citep{KomatsuEtal2003}.

Nevertheless, the three-point correlation function and 
its Fourier-transform pair,
the bispectrum, are complicated objects. In their most general form,
they depend on three vectors, i.e. nine variables.
Even after translational and  rotational  symmetries are taken into account, 
three-point functions depend on the size and shape of 
a triangle, i.e. still three parameters are needed. Exploration 
and visualization of the full configuration space becomes a surprizingly
daunting task,  and most previous studies have been forced to restrict 
their investigation to 
a few hand-picked triangle shapes and sizes, which are not necessarily
representative. 

In the past essentially three distinct parametrizations have been proposed.
\cite{peebles1980} used the parameters $r,u,v$, where the
three sides of the triangles are $r=r_1\le r_2\le r_3$, and $u=r_2/r_1$,
and $v=(r_3-r_2)/r_1$. Alternatively,
\cite{SutoMatsubara1994,deeprange01} 
have used logarithmic bins for the three
sides of a triangle. Then shapes can be uniquely described by a pair of
integers $(b_1-b_3,b_2-b_3)$   
formed from the bin numbers $b_1 \le b_2 \le b_3$.
Finally, several authors parametrized triangles with two sides
and the angle between then, $r_1,r_2,\theta$ \citep[e.g.][]{Fry1994}.

Systematic exploration of the full configuration space, and comparison
of results are somewhat
tedious using any of the above parametrizations. Moreover, the
connection between  real and transform space representations (three-point
function and bispectrum) is somewhat obscure, whichever description
one uses. Hoping to alleviate  these shortcomings, we introduce the
multipole expansion of three-point statistics, which
arguably expresses the rotational symmetries in the most natural way.
The next section presents the definition of the three-point
multipoles and the relationship between the multipoles
in real and Fourier space. 
Section 3 applies these results in
the weakly non-linear regime, bias, and presents predictions
for power law and CDM models.
The final section contains summary and discussion
of the results, including measurement techniques for the
proposed quantities.

\section{Multipole expansion of three-point statistics}

Since the bispectrum can be  parametrized with two
lengths, $k_1,k_2$, and the angle $\theta$ between them,
it is natural to define the following multiple expansion
\begin{equation}
  B(k_1,k_2,\theta)= \sum_l B_l(k_1,k_2)P_l(\cos\theta)\frac{2 l +1}{4 \pi}, 
\end{equation}
where $P_l(\cos\theta)$ is the $l$-th order Legendre-polynomial.
Given the two scalar lengths, $B_l$ can be obtained through
integration $B_l=2\pi \int B P_l d\cos\theta$. 
The multipole expansion of the real space three-point correlation
function, $\xi^3_l$, is defined entirely analogously to $B_l$.

This novel choice of parametrization, while in principle
equivalent to previous choices,  provides
a surprisingly new perspective. It facilitates predictions, measurements,
and their visualizations since it expresses naturally 
the rotational symmetry inherent
in the three-dimensional statistics. To begin with, 
we show that  there is a simple transformation between $B_l$ and $\xi^3_l$.

The relationship of the three-point
correlation function $\xi^3$ to the bispectrum $B$ is a 
Fourier transform:
\begin{equation}
  \xi(r_1,r_2,r_3) = \int\Pi_{i=1}^3d^3k_iB(k_1,k_2,k_3)
  e^{i(k_1x_1+k_2x_2+k_3x_3)}\delta_D(k_1+k_2+k_3),
\end{equation}
$\delta_D$ is the Dirac delta function, and 
temporarily vector notation is used. This is a six dimensional
integral and difficult to treat in practice. 

If the general relation is rewritten in spherical variables in terms of
the above multipole expansion, the exponential can be expanded into
spherical harmonics while the
addition theorem can be used to expand the Legendre polynomial.
The angular dependence then can be integrated using the orthogonality
of spherical functions
yielding the following final result
\begin{equation}
   \xi^3_l(r_1,r_2) = \int\frac{k_1^2}{2 \pi^2}dk_2 \frac{k_2^2}{2 \pi^2} dk_2 
    (-1)^l B_l(k_1,k_2)j_l(k_1r_1)j_l(k_2r_2).
\label{eq:transform}
\end{equation}
The formula is analogous to that of the power spectrum, where a formally
3 dimensional integral can be rendered one dimensional by
$\xi(r) = \int k^2 dk/(2\pi^2)P(k)j_0(k r)$ in the natural
variables.

\section{Theoretical Predictions in the Weakly Non-Linear Regime}

Perturbation theory predicts 
\citep{fry1984a,goroffetal1986,BouchetEtal1995,HivonEtal1995}
that the bispectrum in the weakly non-linear regime is
\begin{equation}
  \left\lbrace\left(\frac{4}{3}+\frac{2}{3}\mu\right)P_0(x)+
  \left(\frac{k_1}{k_2}+\frac{k_2}{k_1}\right)P_1(x)+
  \frac{2}{3}\left(1-\mu\right)P_2(x)\right\rbrace P(k_1)P(k_2)+\rm{perm.} 
\end{equation}
where $x$ is the cosine of the angle between the two wave vectors, $P_l$
are Legendre polynomicals, and
$P(k)$ is the linear power spectrum. The parameter $\mu$ depends
on the expansion history of the universe; its value is
well fitted by $3/7\Omega^{-1/140}$ \citep{KamionkowskiBuchalter1999}.
We have written the above
formula in terms of Legendre polynomials, to make it explicitly
clear that the first permutation depends only on terms up to quadrupole.
The other two permutations, however, will contribute
higher $l$ terms. Typically these higher order terms drop
quickly, although an infrared divergence of the power spectrum
can cause oscillation for isoceles ($k_1=k_2$) configurations only.

In the following we focus on the reduced bispectrum 
$Q=B/(P_1P_2+P_2P_3+P_3P_1)$, for which
the dependence on the amplitude of power spectrum is scaled out,
thus facilitating interpretation.
Analogously to the previous definitions, we
introduce the multipole expansion of the reduced bispectrum, $Q_l(k_1,k_2)$.
Real space version of this quantity $Q_l(r_1,r_2)$ can be introduced 
exactly the same way, however, in what follows, for $Q_l$ will denote
transform space quantity. In the literature the notation $Q_3$ is often used
for the reduced bispectrum: since this paper deals with
three-point statistics only, the index on $Q_l$ denotes multipoles.

Figure 1. plots the multipoles of the reduced bispectrum, $(2l+1)Q_l/4\pi$
for 3 configurations, assuming a power-law power spectrum with
index $n=-1$ for $\Omega=1$. 
Multipoles higher then $l\simeq 2$
appear to be negligible with the possible
exeption of a tiny negative octupole. The right
panel displays the configuration  dependence; $Q_l=Q_l(k_1/k_2)$.
Several values of $k_2$ are plotted, nevertheless they all coincide:
the reduced bispectrum depends only on the ratio of the wave vectors
for a power law initial power spectrum, as expected. 
Therefore, in the weakly non-linear
regime, most of the information contained in the full
configuration space can be displayed in three graphs, 
which show the dependence on the ratio up to quadrupole. 

Most of these results hold for CDM power spectra, with the
exeption that $Q_l$ will depend slightly on the magnitude of the
wave vectors as well.
Next we present predictions for the monopole, dipole, and quadrupole
moment of the reduced bispectrum in CDM models.

\begin{figure}[htb]
\plottwo{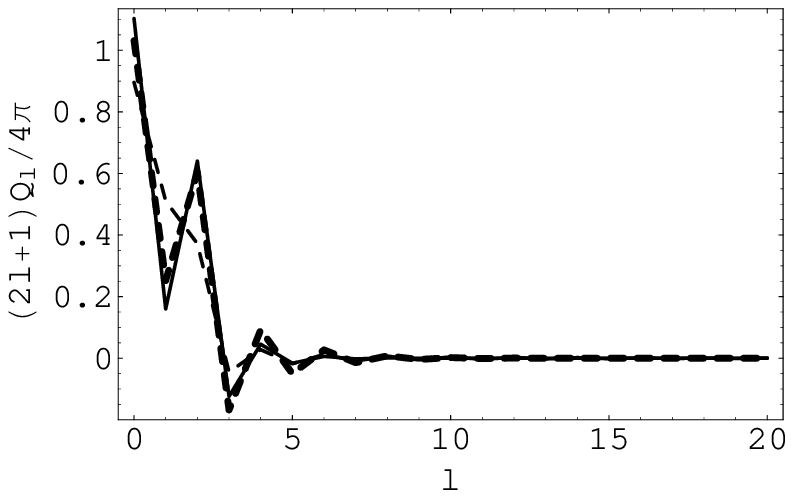}{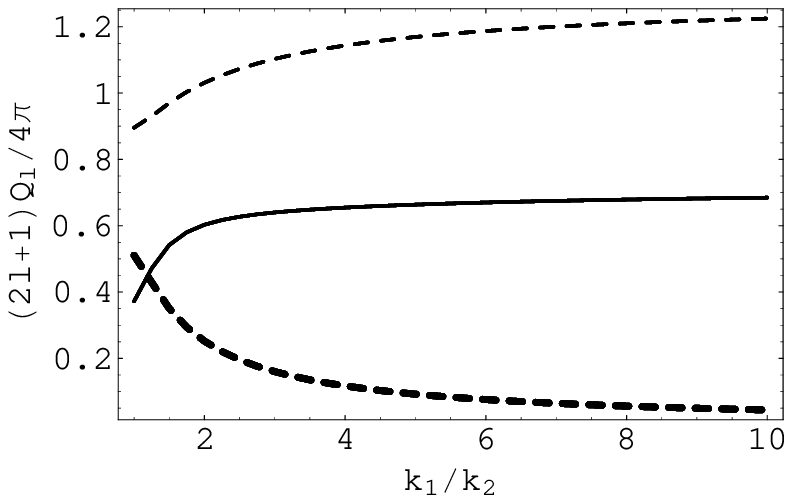}
\caption{Left: The multipole expansion of the reduced bispectrum 
in a power law $n=-1$ model is plotted for
three different configurations of $k_2/k_1$: 
isoceles $1/1$ (thin dash), $2/1$ (thick dash) , 
$4/1$ (solid).
Right: the configuration dependence of the reduced bispectrum, monopole 
(thin dash),  dipole (thick dash),  quadrupole (solid). $k_2$ can be 
fixed at an arbitrary value due to scale invariance of the initial
power spectrum (the actual plot shows three sets of five 
indistinguishable curves calculated
numerically for the same values of $k_2$ as for Figure 2.).
The x axis shows $k_1/k_2$, the shape factor.}
\end{figure} 

We define our concordance
cosmological model based on WMAP \citep{SpergelEtal2003} and
SDSS \citep{TegmarkEtal2003b,PopeEtal2003} with parameters  
$h = 0.685, \Omega = 0.309, \Omega_b = 0.0228/h^2, n=0.966$.
Figure 2 displays results for 
a set of fixed $k_2=0.01-0.05$ as a function of $1 \le k_1/k_2 \le 10$.
As expected, the configuration dependence is not a function
of the ratio only, but  is slowly changing with the magnitude as well.
Still, virtually all information can be summarized in one
plot. The left panel uses a BBKS power spectrum \citep{bardeenetal1986}, 
with no baryonic oscillations.  
The configuration dependence is smooth, qualitatively
similar to the case of the power law power spectrum. The right
panel of uses the same parameters with the
more accurate EH fit power spectrum \citep{EisensteinHu1998};
it features prominent baryonic osciallations.
Baryonic oscillations are also evident on corresponding plots of the
bispectrum $B_l$ as well (not displayed).

\begin{figure}[htb]
\plottwo{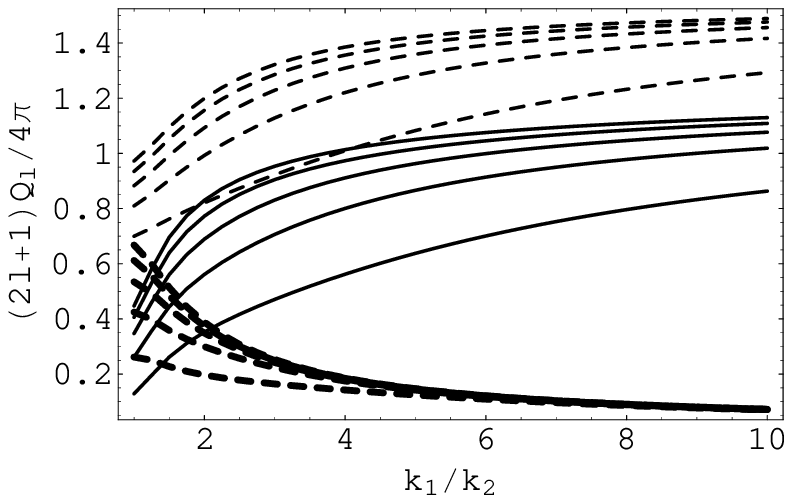}{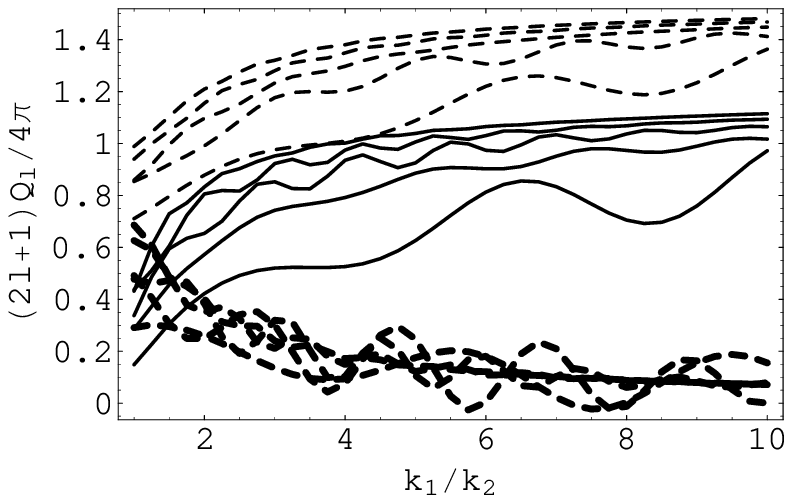}
\caption{Left: 
Left: the configuration dependence of the reduced bispectrum, monopole 
(thin dash),  dipole (thick dash),  quadrupole (solid) is plotted
for a BBKS power spectrum (no baryons). The x axis shows $k_1/k_2$, and the
set of curves display fixed $k_2=0.01,0.02\ldots,0.05$. Right: the
same plot is repeated for EH power spectra, featuring baryonic
oscillations.}
\end{figure} 

\section{Bias}

The above theory can be used to construct estimators which constrain the bias
in the weakly non-linear regime.
Indeed, in the weakly non-linear regime one can expand
the biased field as $f(\delta) = b_0+b_1\delta+b_2/2\delta^2$ 
\citep[e.g.][]{frygaztanaga1993}, where
$\delta$ is the purely gravitational dark matter field. Then the
biased reduced bispectrum transforms as $q = Q/b+b_2/b^2$ 
\citep{Fry1994}, where the lower case denotes the galaxy (measured),
and the upper case the dark matter (theory) values.
It is clear that $b_2$ can only effect the monopole term.
Thus a simple estimator for the bias can be constructed as
\begin{eqnarray}
   b &= \frac{Q_1}{q_1}=\frac{Q_2}{ q_2}\cr
   b_2 &= q_0b^2-Q_0b.
\end{eqnarray}

According to the equations, the  quadrupole to dipole ratio
does not depend on the bias, thus it serves as a novel, useful
test of the underlying assumptions:
a quasi-local perturbative, deterministic bias model and perturbation theory.
Figure 4. shows the dipole to quadrupole ratio for BBKS, 
and EH power spectra, respectively.
The range of $k$'s to be used for bias extraction can be determined from 
contrasting the measurements  with these predictions. Note that
scales where baryon oscillations are prominent are barely accessible
with present redshift surveys. On smaller scales non-linear evolution
is likely to modify these prediction based purely on leading order
perturbation theory (\cite{MeiksinEtal1999}.

\begin{figure}[htb]
\plottwo{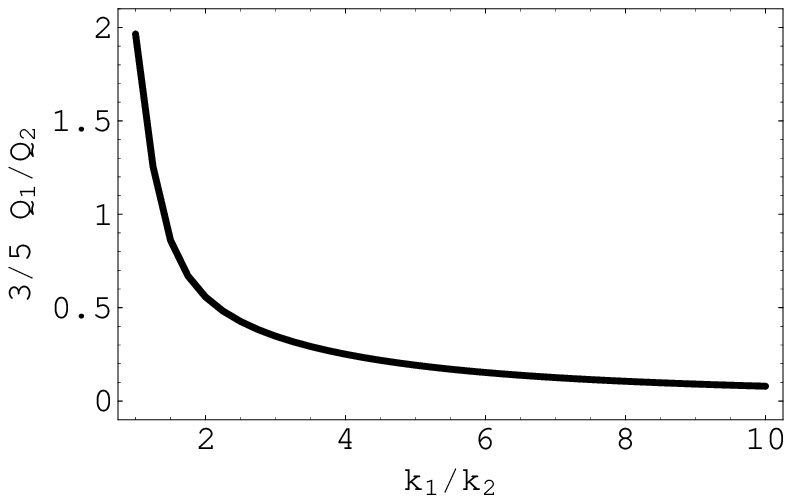}{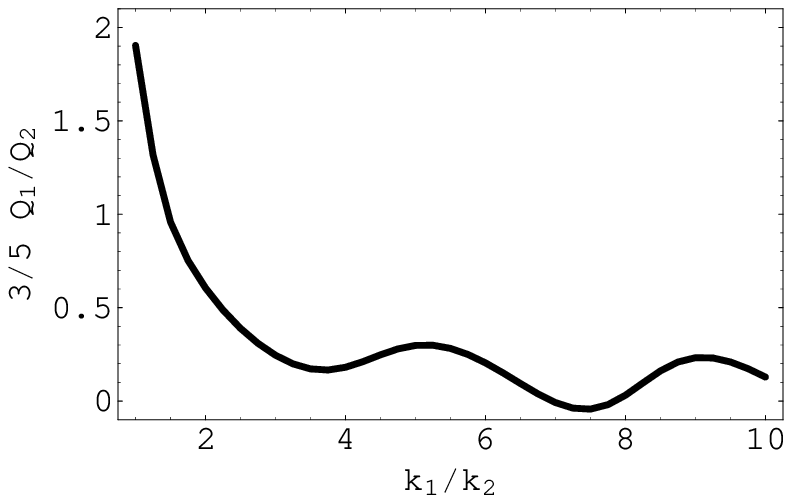}
\caption{Left: the dipole to quadrupole  ratio $3/5 Q_1/Q_2$
is plotted for a BBKS theory (thick solid line) for $k=0.01$. 
The right panel shows the same using the EH fit for comparison,
featuring baryonic oscillations.}
\end{figure} 

\section{Summary and Discussions}

We have introduce the multipole expansion of three-point statistics,
which expresses the inherent  rotational symmetry in the most natural
way.  In particular, the relationship of real and transform space become
entirely transparent via  Hankel transforms.
As a first application, we explored the meaning of the novel quantities
theoretically in the weakly non-linear regime. For the multipole
expansion of the reduced bispectrum, $Q_l(k_1,k_2)$,
most of the information can be compressed into three
functions, monopole, dipole, and quadrupole. From these, we
constructed a novel estimator for the weakly non-linear bias. 
Since the quadrupole to dipole ratio is independent of the bias,
it serves as a novel diagnostic of the underlying
theoretical assumptions. This analogous redshift distortions, where
the quadrupole to monopole ratio plays a similar role.

Our repackaging of three-point statistics
has yielded novel predictions. In particular, the
baryonic oscillations appear to be extremely prominent in 
the presently favoured concordance model, suggesting
that our estimators  will become useful for  main stream parameter
estimation. 

In previous work, a formally similar
parametrization have been used to
describe projection of primordial non-Gaussianities of
the CMB \citep{SpergelGoldberg1999}, and,  in an identical
result, for projecting
the weakly non-linear bispectrum on the sphere \citep{VerdeEtal2002}.
These papers use an intermediate quantity nearly identical to our
definition of $B_l$, thus their results are directly appicable
to the projection of the multipoles on the sphere.

The measurement of $B_l$ merits special consideration. The most immediate
consequence of our theory is that
the $\theta$ dependence of $\xi^3$ or $B_l$ should
be sampled at the roots of Legendre
polynomials of order $l_{max}$, where  $l_{max}$ is the desired
angular resolution. Then Gauss-Legendre integration can yield the 
$\xi^3_l$ or $B_l$.
Starting out in real space, one can use the algorithms
of \cite{MooreEtal2001} to measure $\xi^3$, or, in transform space,
 one can estimate $B$ directly \citep{Scoccimarro2000}.
Either way, and Equation~\ref{eq:transform} transforms one into the other,
which ultimately amounts to a novel {\em edge corrected bispectrum}
estimator. This is significant improvement over all previous
techniques, where no edge correction have been used for the
bispectrum.

A theoretically interesting possibility is to
expand  the estimator of the bispectrum into bipolar spherical harmonics 
$\delta_{k_1}\delta_{k_2}\delta^*_{k_1+k_2}= \sum_{LMl_1l_2}
   b^{LM}_{l_1l_2}(k_1,k_2) (Y_{l_1} \otimes Y_{l_2})_{LM}$
\citep[e.g.][]{VarshalovichEtal1988}. Since the bipolar
spherical harmonics are orthogonal, and 
$(Y_{l} \otimes Y_{l})_{00} \propto P_l$, one can show
that $B_l = b^{00}_{ll} (-1)^l/ \sqrt{2l+1}$. While the bipolar
transformation, especially for low $l$'s, can be calculated
with fast harmonic transform, the simple minded Gaussian quadrature
above is likely to be more practical. Nevertheless, this
sheds light on how $B_l$ is related to the symmetries
on direct product of two spheres $S^2\otimes S^2$.
This is the three-point
generalization of SpICE \cite{spc01},
eSpICE \cite{SzapudiEtal03} algorithms for
fast,  edge corrected estimations of the power spectrum.

To apply these techniques successfully to data in the framework
of high precision cosology, 
it will be necessary to extend our calculations with
redshift distortions \citep[e.g.][]{Kaiser1987,ScoccimarroEtal1999}
taken properly into account.
However, at least one (suboptimal) estimate of the
real space quantities can be obtained from the transverse
power \citep{HamiltonTegmark2002}. The redshift
space bispectrum is parametrized by the five parameters
$B(k_{1,\bot},k_{2,\bot},k_{3,\bot},k_{1,\|},k_{2,\|})$,
with $\bot$ denoting transverse, and $\|$ paralell quantities
with respect to the line of sight in the distant observer
approximation. Then the real space bispectrum can be estimated
from taking $k_{1,\|}\simeq k_{2,\|}\simeq 0$.

In the highly non-linear regime it is expected that 
$l_{max} > 2$ will be needed for full description, since halo
models predict strong features in the configuration dependence
of the three-point correlation function \citep{TakadaJain2003}.
Nevertheless, it is likely that the multipole expansion will
be convenient even in this case.

Several immediate generalizations to the above ideas are in
the works. The above investigations will be applied 
to the halo model to extend the theory for dark matter and galaxies
in any regime. According to \cite{TakadaJain2003}, 
the halo model integrals are 7 dimensional, and
intractable without approximations.
This is no longer the case: with our Equation~\ref{eq:transform}
 the integrals reduce
to three dimensions for each multipole. Generalization of the theory
is fairly straightforward for projected quantities, such as
angular clustering of galaxies, CMB, as well as for vector and tensor
correlations, such as CMB polarizations and weak lensing.
We will be exploring practical implementations of the mentioned
measurement methods, with special attention to practical caveats,
such as binning/regularization and Poisson noise subtraction.
Multipole expansion is quite natural for redshift distortions,
and we aim to generalize our calculation 
for the three-point multipoles, preferably
without making use of the distant observer approximation as in
\cite{SzalayEtal1998}.
This and the above generalizations will be presented 
in subsequent publications.

It is a pleasure to thank Pablo Fosalba, Jun Pan,
and Alex Szalay for stimulating discussions. 
The author was supported by NASA through AISR 
NAG5-11996, and ATP NASA NAG5-12101 as well as by
NSF grants AST02-06243 and ITR 1120201-128440.





\end{document}